\documentclass[aps,prd,showpacs,byrevtex,amsmath,amssymb,superscriptaddress]{revtex4}
\usepackage{dcolumn}
\usepackage{bm}
\usepackage{graphicx,epsfig,verbatim,enumerate}
\usepackage{amssymb}
\usepackage{ifthen}

\begin{document}

\title{Decay of spin-1/2 field around Reissner-Nordstrom black hole }

\markboth{Decay of spin-1/2 field around Reissner-Nordstrom black
hole}{Zhou and Zhu}

\author{
  \firstname{Wei}
  \surname{Zhou}
  }
\affiliation{
        Department of Physics,
        Beijing normal University,
        Beijing 100875,
        China
        }
\author{
  \firstname{Jian-Yang}
  \surname{Zhu}
  }

\affiliation{
        Department of Physics,
        Beijing normal University,
        Beijing 100875,
        China
        }
\affiliation{
        CCAST (World Laboratory),
        Box 8730, Beijing 100080,
        China
        }
\date{\today}

\begin{abstract}
To find what influence the charge of the black hole $Q$ will bring
to the evolution of the quasinormal modes, we calculate the
quasinormal frequencies of the neutrino field (charge $e=0$)
perturbations and those of the massless
Dirac field ($e\neq 0$) perturbations in the RN metric. The influences of $Q$%
, $e$, the momentum quantum number $l$, and the mode number $n$
are discussed. Among the conclusions, the most important one is
that, at the stage of quasinormal ringing, the larger when the
black hole and the field have the same kind of charge ($eQ>0$),
the quasinormal modes of the massless
charged Dirac field decay faster than those of the neutral ones, and when $%
eQ<0$, the massless charged Dirac field decays slower.
\end{abstract}

\pacs{04.70.-s, 04.30.-w} \maketitle

\section{Introduction}

It is well known that black holes and neutron stars have their
characteristic ''sound'': quasinormal modes, which were originally
found to play a prominent role in the gravitational radiation
theoretically and were hoped to be detected in the gravitational
experiments (see\cite {K.Kokkotas,Hans} and references therein for
review). They have been extensively studied recently for their
possible connection with the quantum gravity through an
observation made by Dreyer in Ref. \cite{Dreyer}, where the author
fixed the free parameter $\gamma $ of Loop Quantum Gravity(LQG)
and suggested the gauge group of LQG may be $SO(3)$ rather than
$SU(2)$. This is remarkable\cite{Baez} and surely makes a good
example of utilizing the macroscopic behavior of black holes to
deduce their microscopic nature.

Besides the application in LQG, the quasinormal modes are also
used in many other fields, especially the Anti-de sitter/Conformal
Field Theory (AdS/CFT) correspondence\cite{Birm}. In Ref.
\cite{Horowitz}, from the AdS/CFT correspondence, the quasinormal
modes of the AdS black hole are used to determine the relaxation
time of a field perturbation, i.e., the imaginary part of the
quasinormal frequency is proportional to the inverse of the
damping time of a given mode. Hod and Piran\cite{Hod} considered
the behavior of a charged scalar field perturbation and found
that, its quasinormal modes will dominate the radiation in the
late time evolution for their slower decay than the neutral ones.
Konoplya, in his work\cite {Konoplya}, first investigated a
complex (charged) scalar field perturbation through calculating
its quasinormal modes in the Reissner-Nordstrom (RN), RNAdS and
dialton black holes and found that, on the contrary, the neutral
perturbations will decay slower than the charged ones at the stage
of quasinormal ringing. In this paper, we turn to treat the spin
$1/2$ Dirac particles, in the presence of the RN\ black holes with
charge $Q$. Calculating quasinormal frequencies of the neutrino
field (charge $e=0$) perturbations and those of the massless Dirac
field (charge $e\neq 0$) perturbations in the RN metric, we found
that when the product $eQ$ is positive, the quasinormal modes of
the massless Dirac field decay faster than those of the neutral
ones and when $eQ$ is negative, the perturbations of the massless
Dirac field decay slower than those of the neutral field. The
detailed discussions can be found in the following sections.

In Sec. \ref{Sec.2}, in the RN metric, we will calculate the
quasinormal frequencies of the neutrino field perturbations and
those of the massless Dirac field perturbations. In Sec.
\ref{Sec.3}, with some figures and tables, we show clearly what
influence the charge of the black hole brings to the behavior of
the quasinormal modes, including the energy and the decay rate.
Sec. \ref{Sec.4} is a summary with some suggestions about the
future research.

\section{Quasinormal modes in RN metric}

\label{Sec.2}

We shall calculate the quasinormal frequencies of the massless
Dirac field perturbations and the neutrino field perturbations in
the RN metric
\begin{equation}
ds^2=-\left( 1-\frac{2M}r+\frac{Q^2}{r^2}\right) dt^2+\left( 1-\frac{2M}r+%
\frac{Q^2}{r^2}\right) ^{-1}dr^2+r^2\left( d\theta ^2+\sin
^2\theta d\phi ^2\right) .  \label{metric}
\end{equation}
The wave equation can be written as
\begin{equation}
\lbrack \gamma ^ae_a^{~\mu }(\partial _\mu +\Gamma _\mu +eA_\mu
)]\Psi =0. \label{DiracEq}
\end{equation}
In Eq. (\ref{DiracEq}), $\gamma ^a$ are the Dirac matrices with
the forms
\[
\gamma ^0=\left(
\begin{array}{cc}
-i & 0 \\
0 & i
\end{array}
\right) ,~\gamma ^i=\left(
\begin{array}{cc}
0 & -i\sigma ^i \\
i\sigma ^i & 0
\end{array}
\right) ,~\left( i=1,2,3\right) ,
\]
where $\sigma ^i$ are the Pauli matrices. $e_a^{~\mu }$ is the
inverse of the tetrad $e_\mu ^{~a}$ defined by the spacetime
metric
\[
g_{\mu \nu }=\eta _{ab}e_\mu ^{~a}e_\nu ^{~b},\ \eta
_{ab}=diag(-1,1,1,1).
\]
$\Gamma _\mu $ are the spin connection given by
\[
\Gamma _\mu =\frac 18\left[ \gamma ^a,\gamma ^b\right] e_a^{~\nu
}e_{b\nu ;\mu },\ e_{b\nu ;\mu }=\partial _\mu e_{b\nu }-\Gamma
_{\mu \nu }^ae_{ba},
\]
where $\Gamma _{\mu \nu }^a$ are the Christoffel symbols. And,
$eA_\mu $ is the electromagnetic potential of the black hole.

In the following discussions, we adopt the natural unit, $\hbar
=c=G=4\pi \varepsilon _0=1$. It means that all physical
quantities, such as mass, length, charge, time, energy, and so on,
are dimensionless. The following replacing relations can be used
to restore all of them:

\begin{eqnarray*}
\text{Mass} &:&\text{ }m\rightarrow \left( \frac{c\hbar }G\right) ^{-1/2}m=%
\frac m{m_p},\quad m_p=\sqrt{\frac{c\hbar }G}, \\
\text{Length} &:&\text{ }l\rightarrow \left( \frac{\hbar
G}{c^3}\right)
^{-1/2}l=\frac l{l_p},\quad l_p=\sqrt{\frac{\hbar G}{c^3}}, \\
\text{Time} &:&\text{ }t\rightarrow \left( \frac{\hbar
G}{c^5}\right)
^{-1/2}t=\frac t{t_p},\quad t_p=\sqrt{\frac{\hbar G}{c^5}}, \\
\text{Charge} &:&\text{ }e\rightarrow \left( 4\pi \varepsilon
_0c\hbar
\right) ^{-1/2}e=\frac e{e_p},\quad e_p=\sqrt{4\pi \varepsilon _0c\hbar }, \\
\text{Energy} &:&\text{ }E\rightarrow \left( \frac{c^5\hbar
}G\right) ^{-1/2}E=\frac E{E_p},\quad E_p=\sqrt{\frac{c^5\hbar
}G}.
\end{eqnarray*}

For the RN spacetime, the electromagnetic potential $eA_\mu $ can
be written as
\begin{equation}
eA_\mu =\left( \frac{eQ}r,0,0,0\right) ,  \label{A}
\end{equation}
where $Q$ is the charge of the RN black hole, and $e$ is zero for
neutrinos and nonzero for Dirac particles. In the following we
shall see that we can just consider the sign of the product $eQ$
without taking into consideration the sign of each quantity
respectively. Therefore, for convenience we shall take $Q$ to be
always positive and only switch the sign of $e$, when we are
studying the influence of the charge of the black hole on the
quasinormal frequencies of the Dirac field perturbations.

For the RN spacetime, we can select the tetrad $e_\mu ^{~a}$ as
\begin{equation}
e_\mu ^{~a}=diag\left( \left( 1-\frac{2M}r+\frac{Q^2}{r^2}\right)
^{1/2},\left( 1-\frac{2M}r+\frac{Q^2}{r^2}\right) ^{-1/2},r,r\sin
\theta \right) .  \label{tetrad}
\end{equation}
Using the ansatz \cite{Cho},
\begin{equation}
\Phi (t,r,\theta ,\phi )=\left(
\begin{array}{l}
\frac{iG^{(\pm )}(r)}r\varphi _{jm}^{(\pm )}(\theta ,\phi ) \\
\frac{F^{(\pm )}(r)}r\varphi _{jm}^{(\pm )}(\theta ,\phi )
\end{array}
\right) e^{-iEt},  \label{ansatz}
\end{equation}
where for $j=l+1/2$,
\[
\varphi _{jm}^{(+)}=\left(
\begin{array}{l}
\sqrt{\frac{l+1/2+m}{2l+1}}Y_l^{m-1/2} \\
\sqrt{\frac{l+1/2-m}{2l+1}}Y_l^{m+1/2}
\end{array}
\right) ,
\]
and for $j=l-1/2$,
\[
\varphi _{jm}^{(-)}=\left(
\begin{array}{l}
\sqrt{\frac{l+1/2-m}{2l+1}}Y_l^{m-1/2} \\
-\sqrt{\frac{l+1/2+m}{2l+1}}Y_l^{m+1/2}
\end{array}
\right) ,
\]
we can get the radial part of the wave equation
\begin{equation}
\left( -\frac{d^2}{dr_{*}^2}+V_1\right) F=\omega ^2F,
\label{RadialEq-1}
\end{equation}
and
\begin{equation}
\left( -\frac{d^2}{dr_{*}^2}+V_2\right) G=\omega ^2G,
\label{RadialEq-2}
\end{equation}
where
\[
\frac{dr_{*}}{dr}=\left( 1-\frac{2M}r+\frac{Q^2}{r^2}\right)
^{-1}.
\]
$V_{1,2}$ can be expressed as
\begin{equation}
V_{1,2}(r,k)=\pm \frac{dW}{dr_{*}}+W^2,  \label{V}
\end{equation}
with
\begin{equation}
W=\left( 1-\frac{2M}r+\frac{Q^2}{r^2}\right) ^{\frac 12}\frac{\mid k\mid }r-%
\frac{eQ}r,  \label{W}
\end{equation}
where $k$ equals $l$ in the ($+$) case and $-\left( l+1\right) $ in the ($-$%
) case, and the momentum quantum number $l$ is a non-negative
integer. One may notice that $V_1$ and $V_2$ are supersymmetric
partners of the superpotential $W$\cite{super}, and they will give
the same quasinormal frequencies. So, in the following work, we
only discuss the $V_1(+)$ case where the momentum quantum number
$l$ equals $k$.

\begin{figure}
\epsfig{
         file=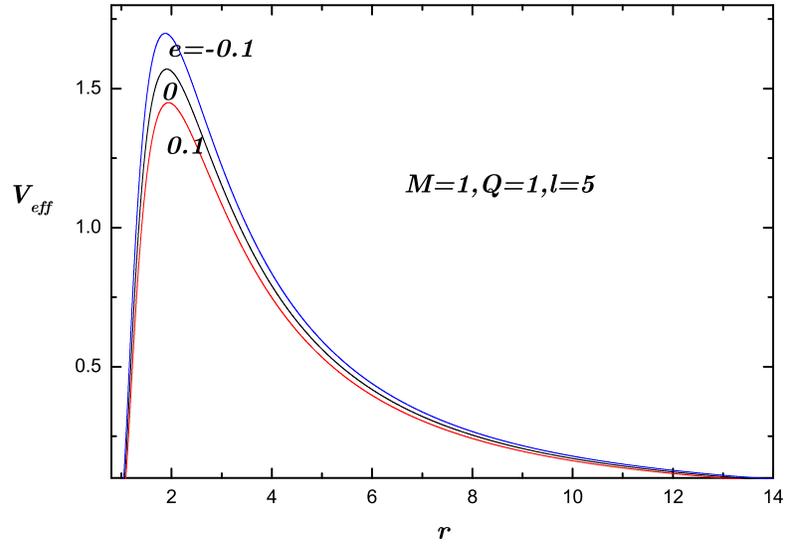, width=0.65\textwidth, angle=0, clip=
         }
\caption{The effective potential barrier of $e=0$, $\pm 0.1$ with $l=5$, $M=1$%
, $Q=1$.}
\label{Fig.1}
\end{figure}

\begin{figure}
\epsfig{
        file=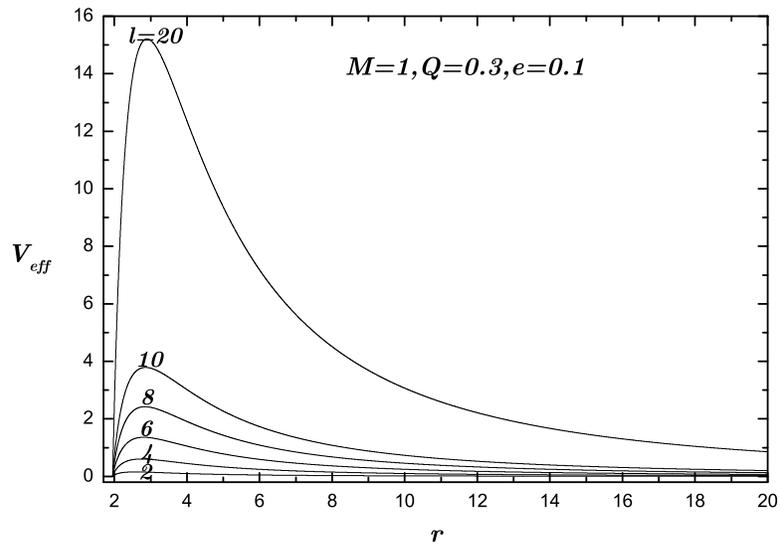, width=0.65\textwidth, angle=0, clip=
         }
\caption{The variation of the potential barrier for the massless
Dirac field with $r$ at $M=1$, $Q=0.3$, $e=0.1$.} \label{Fig.2}
\end{figure}

Fig. \ref{Fig.1} shows the potential barrier of the three cases
mentioned above with the maximal values
\[
V_{\max }(l=5,e=0.1,Q=1)=1.441,
\]
\[
V_{\max }(l=5,e=-0.1,Q=1)=1.689,
\]
\[
V_{\max }(l=5,e=0,Q=1)=1.545,
\]
which can be derived from Eq. (\ref{V}). And from Fig.
\ref{Fig.2}, we can see the peak of the potential barrier
increases with $l$ and its position $r$ approaches a limit.

To evaluate the quasinormal frequencies, we adopt the third-order
WKB approximation, which was originally proposed by Schutz, Iyer
and Will\cite {Schutz} and recently developed by Konoplya to the
sixth-order beyond the eikonal approximation\cite{Konoplya1}. This
method has wide applications in
many black hole cases, and has high accuracy for the low-lying modes with $%
n<l$, where $n$ and $l$ are the mode number and the momentum
quantum number respectively. From Ref. \cite{Iyer}, the expression
of the quasinormal frequencies is
\[
\omega ^2=\left[ V_0+\left( -2V_0^{\prime \prime }\right)
^{1/2}\Lambda \right] -i\left( n+\frac 12\right) \left(
-2V_0^{\prime \prime }\right) ^{1/2}\left( 1+\Omega \right)
\]
where $\Lambda $ and $\Omega $ are the second and third order WKB
correction terms
\begin{eqnarray}
\Lambda &=&\frac 1{(-2V_0^{\prime \prime })^{1/2}}\left\{ \frac
18\left( \frac{V_0^{(4)}}{V_0^{^{\prime \prime }}}\right) \left(
\frac 14+\alpha
^2\right) -\frac 1{288}\left( \frac{V_0^{^{\prime \prime \prime }}}{%
V_0^{^{\prime \prime }}}\right) ^2\left( 7+60\alpha ^2\right)
\right\}
\nonumber \\
\Omega &=&\frac 1{(-2V_0^{\prime \prime })}\left\{ \frac 5{6912}\left( \frac{%
V_0^{\prime \prime \prime }}{V_0^{\prime \prime }}\right) ^4\left(
77+188\alpha ^2\right) -\frac 1{384}\left( \frac{V_0^{\prime
\prime \prime 2}V_0^{(4)}}{V_0^{\prime \prime 3}}\right) \left(
51+100\alpha ^2\right)
\right.  \nonumber \\
&&\ \ \ +\frac 1{2304}\left( \frac{V_0^{(4)}}{V_0^{\prime \prime
}}\right) ^2(67+68\alpha ^2)+\frac 1{288}\left( \frac{V_0^{\prime
\prime \prime }V_0^{(5)}}{V_0^{\prime \prime 2}}\right) \left(
19+28\alpha ^2\right)
\nonumber \\
&&\ \ \ \left. -\frac 1{288}\left( \frac{V_0^{(6)}}{V_0^{\prime \prime }}%
\right) \left( 5+4\alpha ^2\right) \right\} ,
\end{eqnarray}
with $\alpha ^2=\left( n+1/2\right) ^2$. Here, $V_0$ is the
maximal value of
the potential, and $V_0^{\prime \prime }$, $V_0^{\prime \prime \prime }$, $%
V_0^{(i)}$ are the second, third and $i$th order derivatives of
$V_0$, with respect to $r_{*}$.

Subsituting the effective potential of Eq. (\ref{V}) into the
above formula,
we can get the quasinormal frequencies for the three classes: $e=0$, $e<0$, $%
e>0$. To get numerical results, the parameters $Q$, $M$, and $e$
should be
fixed. Without loss of generality, we take $M=1$, $Q=0$, $0.3$, $0.6$, $0.9$%
, $0.99$, $1$ and $e=0$,$~\pm 0.1$, $\pm 0.3$. The results are
listed in the Tables \ref{table1}, \ref{table2}, \ref{table3} and
\ref{table4}, respectively.

\section{Discussions and Conclusions}

\label{Sec.3}

As the main concern of this paper, we will discuss the influence
of the charge of the RN black hole on the behavior of quasinormal
modes in two aspects: how do the quasinormal modes vary with
respect to $e$ (the charge of the particle) and $Q$ (the charge of
the black hole). And following that
we will also discuss the momentum quantum number $l$ and the mode number $n$%
. Thus, this section can be separated into four parts.

From Eq. (\ref{W}), we know that we can just consider the sign of
the product $eQ$ without taking into account the sign of each
quantity respectively. As is mentioned above, we take $Q$ to be
always positive, and only switch the sign before $e$. Thus, before
we move into the discussions, we have to emphasize that positive
values of $e$ actually means that the product $eQ>0$, and {\it
vice versa}.

In the following discussions of the quasinormal modes, we focus on
the real part and the imaginary part respectively. The real part
gives us information of the energy of the emitted particles and
from the imaginary part we can find out the damping time.

$(1)$ {\it The influence of the particle charge} $e$: From Table \ref{table1}%
, we can see how the quasinormal modes behave when the field
changes from neutral to charged cases.

\begin{figure}
\epsfig{
         file=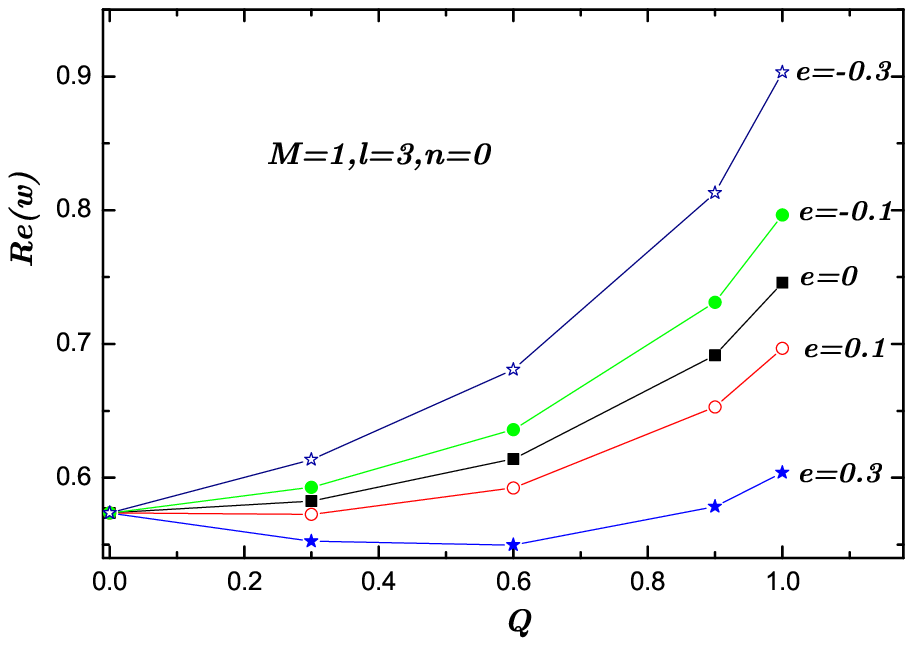, width=0.65\textwidth, angle=0, clip=
         }
\caption{Real part of Quasinormal frequencies of $e=0$, $\pm 0.1$,
$\pm 0.3$ with $Q$ at $M=1$, $l=3$, $n=0$. } \label{Fig.3}
\end{figure}

$(1.1)$ First we study the real part, $Re(w)$: As an example, we
study the case, $Q=0.3$, $l=1$, $n=0$. Here three values $0.1701$,
$0.1799$, $0.1900$
correspond to $e=0.1$, $e=0$, $e=-0.1$, respectively. We can see that $%
Re(w)_{e>0}<Re(w)_{e=0}<Re(w)_{e<0}$. Actually, this result is
valid for all cases in Table \ref{table1}. In Table \ref{table2}
we can further find out the influence of the absolute value of
$e$, and a monotonic behavior is
revealed in Fig. \ref{Fig.3}, i.e., for each given value of $Q$, $%
\mathop{\rm Re}%
\left( \omega \right) $ increases as $e$ decreases.

\begin{figure}
\epsfig{
         file=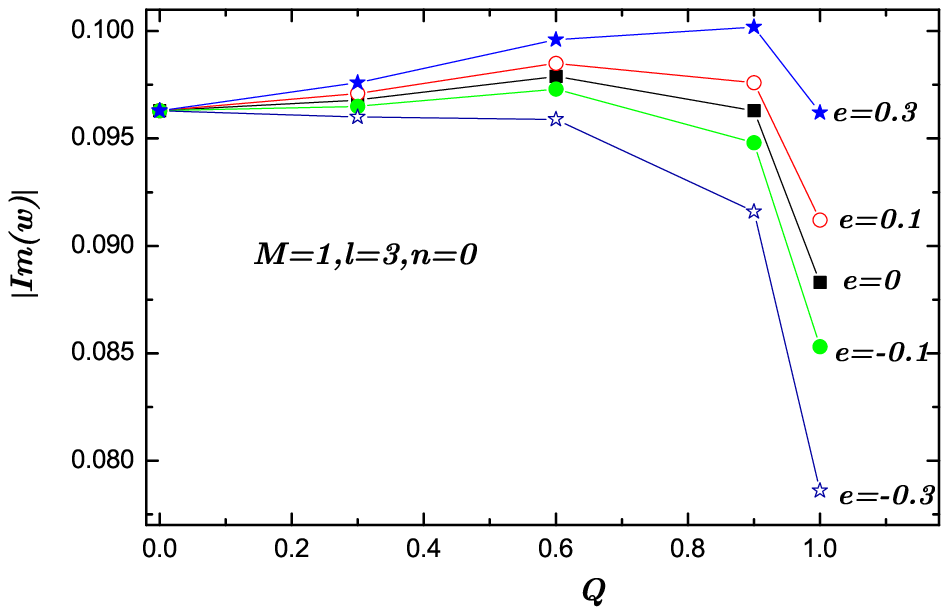, width=0.65\textwidth, angle=0, clip=
         }
\caption{Absolute value of imaginary part of Quasinormal frequencies of $e=0$%
, $\pm 0.1$, $\pm 0.3$ with $Q$ at $M=1$, $l=3$, $n=0$.}
\label{Fig.4}
\end{figure}

$(1.2)$ Then we turn to the imaginary part, $%
\mathop{\rm Im}%
\left( \omega \right) $, of the quasinormal modes. Similarly,
first we consider the sign of $e$. In Table \ref{table1}, we can
see that when $e$ is $0.1$, the absolute values of the imaginary
part of quasinormal frequencies are bigger than those of the
neutrino field and when $e$ is $-0.1$, the
situation is to the contrary. If we take into consideration the value of $e$%
, as is shown in Table \ref{table2} and Fig. \ref{Fig.4}, we also
observe a
monotonic behavior, i.e., the absolute value of $%
\mathop{\rm Im}%
\left( \omega \right) $ increases as $e$ increases.

If we bear in mind that the sign of $e$ actually denotes that of
the product $eQ$, we can summarize the above conclusions as the
following: with larger values of $e$, the particles have smaller
energy and they decay faster.

$(2)$ {\it The influence of the black hole charge} $Q$: In recent
studies reported in Ref. \cite{Konoplya}, the author investigated
the influence of the black hole charge on the complex scalar field
perturbation (charged) in the RN spacetime. It is shown clearly in
the figures that the real part of the quasinormal modes of both
neutral and charged scalar field grows as $Q$ increases, and the
latter grows more rapidly. On the other hand, the imaginary part
of a given ''charged mode'' approaches the neutral one in the
extremal limit $(Q=1)$. In this paper, following the suggestion of
Ref. \cite {Konoplya}, we calculate the higher multipole number
perturbations of the Dirac field with $l=3$, $4$, $5$ and $n=0$,
and the results can be found in Table \ref{table2}.

$(2.1)$ The real part, $%
\mathop{\rm Re}%
\left( \omega \right) $: From Fig. \ref{Fig.3} we can see it also
increases
with $Q$ and another character: compared with the neutrino field, $%
\mathop{\rm Re}%
\left( \omega \right) $ of the perturbations charged $0.1$ and
$0.3$ are smaller in both the value and the increasing speed with
respect to $Q$, however things are different in the cases $e=-0.1$
and $e=-0.3$ which shows the influence from the $Q$ mentioned
above: making the perturbations weaker in the case $e>0$ and
stronger in the case $e<0.$

$(2.2)$ The imaginary part, $%
\mathop{\rm Im}%
\left( \omega \right) $: From Fig. \ref{Fig.4} and Table
\ref{table2}, we find although no coincidence of the neutrino and
the charged Dirac field perturbations in the extremal limit
exists, there is a turning point separating a monotonically
increasing region and a decreasing region (note that there is a
similar point for the scalar fields, see Fig. 2 of Ref. \cite
{Konoplya}). The explanation of this fact is still unclear but
some further work may be helpful, such as finding the relationship
between the position of the turning point and the value of $Q$,
for different fields.

\begin{figure}
\epsfig{
         file=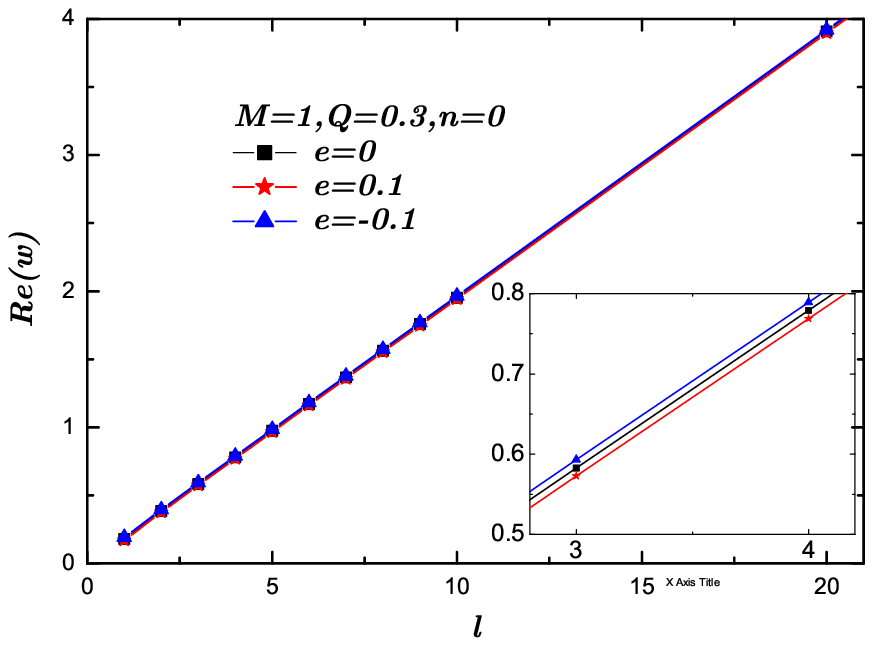, width=0.65\textwidth, angle=0, clip=
         }
\caption{Real part of Quasinormal frequencies of $e=0$, $\pm 0.1$
with $l$ at $M=1$, $Q=0.3$, $n=0$.} \label{Fig.5}
\end{figure}

\begin{figure}
\epsfig{
         file=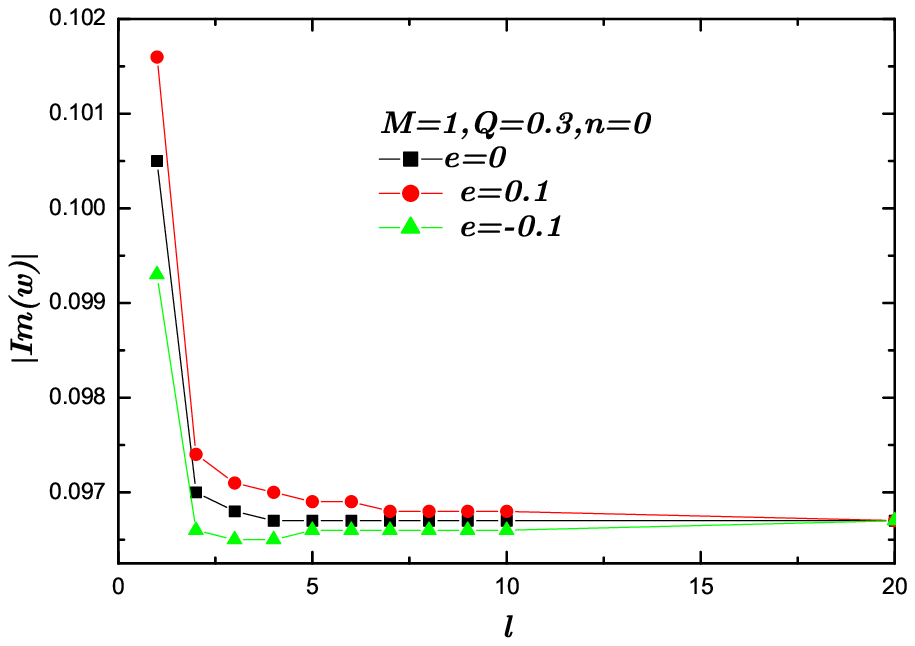, width=0.65\textwidth, angle=0, clip=
         }
\caption{Absolute value of imaginary part of Quasinormal frequencies of $e=0$%
, $\pm 0.1$ with $l$ at $M=1$, $Q=0.3$, $n=0$.}
\label{Fig.6}
\end{figure}

$(3)$ {\it The influence of }$l$: It can be revealed by tuning the value of $%
l$ with the other parameters fixed. As an example, taking $n=0$, $Q=0.3$, $%
e=0$, $\pm 0.1$, from Table \ref{table3} and Fig. \ref{Fig.5}, we
can see
that $%
\mathop{\rm Re}%
\left( \omega \right) $ keeps growing with $l$ at the speed of
$0.2$, which shows the contribution of the angular momentum to the
energy of the emitted particles. As is already discussed above,
from Fig. \ref{Fig.5}, we can also observe the almost constant
difference induced by the values of $e$. From
Fig. \ref{Fig.6} we find that $%
\mathop{\rm Im}%
\left( \omega \right) $ drops sharply as soon as $l$ increases
from zero and then it approaches $0.0967$ asymptotically, a value
independent of $e$. This is not difficult to explain. When the
particle's angular momentum is big enough, it will hide the
influence brought by the different charges. It is also verified by
the potential barrier: when $Q$ is $0.3$ and $l$ is large
enough, the maximal potential value with different values of $e$ is always $%
0.03819l^2$.

\begin{figure}
\epsfig{
         file=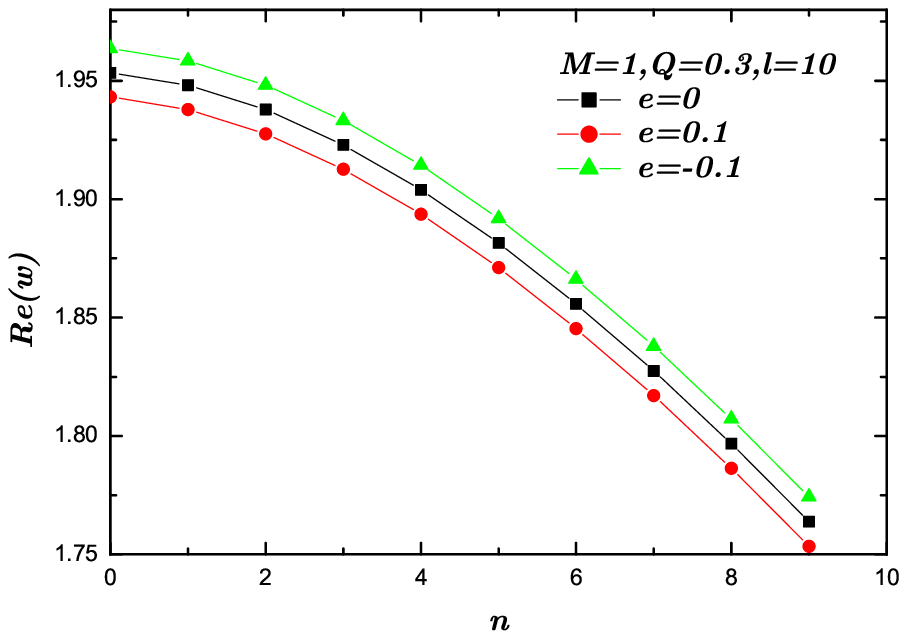, width=0.65\textwidth, angle=0, clip=
         }
\caption{Real part of Quasinormal frequencies of $e=0$, $\pm 0.1$
with $n$ at $M=1$, $Q=0.3$, $l=10$.} \label{Fig.7}
\end{figure}

\begin{figure}
\epsfig{
         file=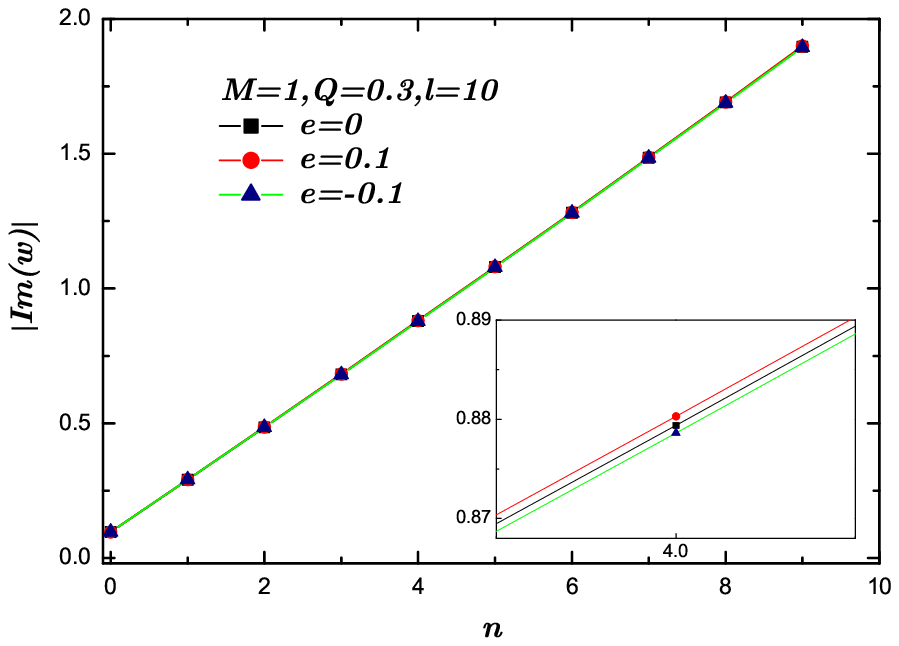, width=0.65\textwidth, angle=0, clip=
         }
\caption{Absolute value of imaginary part of Quasinormal
frequencies of $e=0$, $\pm 0.1$ with $n$ at $M=1$, $Q=0.3$,
$l=10$.} \label{Fig.8}
\end{figure}

$(4)$ {\it The influence of }$n$: A related discussion can be
found in Ref. \cite{Cho}. It can be revealed by tuning the value
of $n$ with the other parameters fixed. As an example, taking
$l=10$, $Q=0.3$, $e=0$, $\pm 0.1$, we can see from Fig.
\ref{Fig.7}, \ref{Fig.8}, and Table \ref{table4} that
when $n$ increases, $%
\mathop{\rm Re}%
\left( \omega \right) $ decreases and $\left|
\mathop{\rm Im}%
\left( \omega \right) \right| $ increases. This clearly shows that
the low-lying modes are of longer relaxation time in not only the
schwarzschild spacetime\cite{Cho} but also in the RN spacetime,
and thus are more important for the description of the field
evolution around black holes.

\section{Summary and suggestions}

\label{Sec.4}

To summarize, in this paper the quasinormal modes of the spin
$1/2$ field are discussed at the presence of the
Reissner-Nordstrom black hole. We study both the real part and the
imaginary part of the quasinormal frequencies with different
values of the particle charge $e$, the black hole charge $Q$, the
momentum quantum number $l$, and the mode number $n$. A series of
conclusions have been reached, among which the most important one
is that, at the stage of quasinormal ringing, when the black hole
and the field have the same kind of charge ($eQ>0$), the
quasinormal modes of the massless
charged Dirac field decay faster than those of the neutral ones, and when $%
eQ<0$, the massless charged Dirac field decay slower.

Concerning the same problem, the decay of the spin $1/2$ field
around black holes, there are still a number of interesting works
to be completed. One work is to consider the massive Dirac field
and focus on the effect of mass by comparing it with the massless
field studied here. The calculation would be more complicated
because the mass of the particle will appear in the Dirac equation
and the effective potential. But we believe that, following the
method in Ref. \cite{Cho}, where the quasinormal frequencies of
the (massless and massive) Dirac field perturbations with the
Schwarzschild black hole are calculated, the massive case with
charged black holes can also be worked out.

One may also be interested in comparing the present results in the
RN spacetime with those in some other charged black holes such as
the RNAdS and dilaton black holes. An especially intriguing case
is the Kerr black hole, with non zero angular momentum. By
comparing the quasinormal frequencies at the presence of
non-rotating black holes and rotating black holes, like the Kerr
black hole\cite{Kerr}, we may understand what influence the black
hole rotation brings to the quasinormal modes.

\acknowledgments

This work was supported by the National Natural Science Foundation
of China under Grant No. 10075025. W. Zhou. thanks Dr. H. B. Zhang
and. Dr. Z. J. Cao for their zealous help and valuable
discussions.

\begin{table}
\caption{The quasinirmal frequencies for RN black hole, $l=3$,
$4$, $5$, $n=0$, $e=0$, $\pm 0.1$, $\pm 0.3$, $M=1$, $Q=0$, $0.3$,
$0.6$, $0.9$, $0.99$.}
\begin{ruledtabular}
\begin{tabular}{ccccccccccc}
\multicolumn{1}{c}{$l$} &\multicolumn{1}{c}{$n$}
&\multicolumn{1}{c}{$e$} &\multicolumn{2}{c}{$Q=0.3$}
&\multicolumn{2}{c}{$Q=0.6$} &\multicolumn{2}{c}{$Q=0.9$}
&\multicolumn{2}{c}{$Q=0.99$}\\
$$&$$&$$&$Re(w)$&$Im(w)$&$Re(w)$&$Im(w)$&$Re(w)$&$Im(w)$&$Re(w)$&$Im(w)$
\\
\hline 1
&0&0.1&0.1701&-0.1016&0.1712&-0.1034&0.1835&-0.1018&0.1879&-0.0982\\
&&0&0.1799&-0.1005&0.1919&-0.1012&0.2207&-0.0979&0.2336&-0.0902\\
&&-0.1&0.1900&-0.0993&0.2138&-0.0987&0.2610&-0.0929&0.2849&-0.0808\\

2
 &0&0.1&0.3746&-0.0974&0.3844&-0.0990&0.4197&-0.0984&0.4418&-0.0936\\
 &&0&0.3847&-0.0970&0.4059&-0.0981&0.4581&-0.0963&0.4891&-0.0898\\
 &&-0.1&0.3949&-0.0966&0.4279&-0.0972&0.4978&-0.0941&0.5387&-0.0856\\
 &1&0.1&0.3499&-0.3016&0.3614&-0.3059&0.4008&-0.3017&0.4205&-0.2863\\
 &&0&0.3602&-0.2999&0.3834&-0.3023&0.4396&-0.2944&0.4668&-0.2740\\
 &&-0.1&0.3706&-0.2982&0.4057&-0.2987&0.4795&-0.2868&0.5152&-0.2609\\

3
&0&0.1&0.5726&-0.0971&0.5923&-0.0985&0.6528&-0.0976&0.6912&-0.0924\\
 &&0&0.5827&-0.0968&0.6140&-0.0979&0.6915&-0.0963&0.7390&-0.0899\\
 &&-0.1&0.5929&-0.9651&0.6360&-0.0973&0.7311&-0.0948&0.7882&-0.0872\\
 &1&0.1&0.5552&-0.2953&0.5762&-0.2994&0.6397&-0.2959&0.6716&-0.2799\\
 &&0&0.5655&-0.2943&0.5982&-0.2973&0.6787&-0.2913&0.7242&-0.2715\\
 &&-0.1&0.5759&-0.2933&0.6204&-0.2952&0.7185&-0.2865&0.7730&-0.2632\\
 &2&0.1&0.5268&-0.5010&0.5498&-0.5076&0.6178&-0.4997&0.6454&-0.4802\\
 &&0&0.5372&-0.4992&0.5719&-0.5037&0.6566&-0.4914&0.6971&-0.4578\\
 &&-0.1&0.5476&-0.4974&0.5942&-0.4997&0.6962&-0.4830&0.7445&-0.4436\\

\end{tabular}
\end{ruledtabular}
\label{table1}
\end{table}

\begin{table}
\caption{The quasinirmal frequencies for RN black hole, $l=3$,
$4$, $5$, $n=0$, $e=0$, $\pm 0.1$, $\pm 0.3$,  $M=1$, $Q=0$,
$0.3$, $0.6$, $0.9$.}
\begin{ruledtabular}
\begin{tabular}{cccccccccc}
\multicolumn{1}{c}{$l$} &\multicolumn{1}{c}{$e$}
&\multicolumn{2}{c}{$Q=0$} &\multicolumn{2}{c}{$Q=0.3$}
&\multicolumn{2}{c}{$Q=0.6$}
&\multicolumn{2}{c}{$Q=0.9$}\\
$$&$$&$Re(w)$&$Im(w)$&$Re(w)$&$Im(w)$&$Re(w)$&$Im(w)$&$Re(w)$&$Im(w)$
\\
\hline 3
 &0.3&&&0.5525&-0.0976&0.5498&-0.0996&0.5783&-0.1002\\
 &0.1&&&0.5726&-0.0971&0.5923&-0.0985&0.6528&-0.0976\\
 &0&0.5737&-0.0963&0.5827&-0.0968&0.6140&-0.0979&0.6915&-0.0963\\
 &-0.1&&&0.5929&-0.0965&0.6360&-0.0973&0.7311&-0.0948\\
 &-0.3&&&0.6135&-0.0960&0.6808&-0.0959&0.8130&-0.0916\\

4
 &0.3&&&0.7488&-0.0976&0.7561&-0.0992&0.8095&-0.0994\\
 &0.1&&&0.7690&-0.0970&0.7991&-0.0983&0.8850&-0.0974\\
 &0&0.7672&-0.0963&0.7792&-0.0967&0.8208&-0.0979&0.9239&-0.0963\\
 &-0.1&&&0.7894&-0.0965&0.8428&-0.0974&0.9634&-0.0952\\
 &-0.3&&&0.8099&-0.0961&0.8873&-0.0964&1.0444&-0.0928\\

5
&0.3&&&0.9448&-0.0972&0.9623&-0.0990&1.0408&-0.0988\\
&0.1&&&0.9650&-0.0969&1.0053&-0.0982&1.1169&-0.0972\\
&0&0.9602&-0.0963&0.9752&-0.0967&1.0272&-0.0979&1.1559&-0.0963\\
&-0.1&&&0.9854&-0.0966&1.0491&-0.0975&1.1953&-0.0954\\
&-0.3&&&1.0059&-0.0962&1.0936&-0.0967&1.2759&-0.0935\\

\end{tabular}
\end{ruledtabular}
\label{table2}
\end{table}

\begin{table}
\caption{ The quasinirmal frequencies for RN black hole, $l=1$,
..., $5$, $10$, $20$, $n=0$, $Q=0.3$, $e=0$, $\pm 0.1$.}
\begin{ruledtabular}
\begin{tabular}{ccccccccc}
$$ &$e$&$l=1$&$l=2$&$l=3$&$l=4$&$l=5$&$l=10$&$l=20$\\
\hline
$Re(w)$& 0&0.1799&0.3847&0.5827&0.7792&0.9752&1.9534&3.9082\\&0.1&0.1701&0.3746&0.5726&0.7690&0.9650&1.9432&3.8980\\&-0.1&0.1900&0.3949&0.5929&0.7894&0.9854&1.9636&3.9184\\
$Im(w)$& 0&-0.1005&-0.0970&-0.0968&-0.0967&-0.0967&-0.0967&-0.0967\\&0.1&-0.1016&-0.0974&-0.0971&-0.0970&-0.0969&-0.0968&-0.0967\\&-0.1&-0.0993&-0.0966&-0.9651&-0.0965&-0.0966&-0.0966&-0.0967\\
\end{tabular}
\end{ruledtabular}
\label{table3}
\end{table}

\begin{table}
\caption{The quasinirmal frequencies for RN black hole,
$n=0,...,7$, $l=8$, $Q=0.3$, $e=0$, $\pm 0.1$.}
\begin{ruledtabular}
\begin{tabular}{cccccccccc}
$$ &$e$&$n=0$&$n=1$&$n=2$&$n=3$&$n=4$&$n=5$&$n=6$&$n=7$\\
\hline
$Re(w)$& 0&1.9534&1.9481&1.9379&1.9230&1.9040&1.8815&1.8558&1.8257\\&0.1&1.9432&1.9379&1.9276&1.9127&1.8937&1.8711&1.8454&1.8171\\&-0.1&1.9636&1.9584&1.9481&1.9333&1.9144&1.8918&1.8662&1.8379\\
$Im(w)$& 0&-0.0967&-0.2904&-0.4852&-0.6814&-0.8794&-1.0795&-1.2814&-1.4852\\&0.1&-0.0968&-0.2907&-0.4856&-0.6820&-0.8803&-1.0806&-1.2827&-1.4867\\&-0.1&-0.0966&-0.2902&-0.4847&-0.6807&-0.8786&-1.0784&-1.2801&-1.4837\\
\end{tabular}
\end{ruledtabular}
\label{table4}
\end{table}

\end{document}